\documentclass[aps,preprint,letterpaper,amsmath,amssymb]{revtex4}
\usepackage{graphicx}
\usepackage{dcolumn}
\usepackage{bm}
\begin{document}

\def\vri{\vec{r}_{i}} \def\vrj{\vec{r}_{j}} \def\rij{r_{ij}}
\def\vrij{\vec{r}_{ij}} \def\drij{\hat{r}_{ij}}
\def\vdr{\delta\vec{r}} \def\dr{\delta{r}} \def\s{\hat{s}}

\title{
  A first-order phase transition at the random close packing of hard
  spheres}

\author{Yuliang Jin and Hern\'an A. Makse}

\affiliation {Levich Institute and Physics Department, City College of
  New York, New York, NY 10031, US}


\begin{abstract} {\bf Randomly packing spheres of equal size into a
    container consistently results in a static configuration with a
    density of $\sim$64\%.  The ubiquity of random close packing (RCP)
    rather than the optimal crystalline array at 74\% begs the
    question of the physical law behind this empirically deduced
    state. Indeed, there is no signature of any macroscopic quantity
    with a discontinuity associated with the observed packing
    limit. Here we show that RCP can be interpreted as a manifestation
    of a thermodynamic singularity, which defines it as the ``freezing
    point'' in a first-order phase transition between ordered and
    disordered packing phases. Despite the athermal nature of granular
    matter, we show the thermodynamic character of the transition in
    that it is accompanied by sharp discontinuities in volume and
    entropy. This occurs at a critical compactivity, which is the
    intensive variable that plays the role of temperature in granular
    matter.  Our results predict the experimental conditions necessary
    for the formation of a jammed crystal by calculating an analogue
    of the ``entropy of fusion". This approach is useful since it maps
    out-of-equilibrium problems in complex systems onto simpler
    established frameworks in statistical mechanics.}
\end{abstract}



\maketitle

\clearpage




Since the time of Kepler it is thought that the most efficient packing
of monodisperse spherical grains is the face centered cubic (FCC)
arrangement with a density of 74 \% \cite{aste-book}.  Thus, we might
expect that spherical particles will tend to optimize the space they
occupy by crystallizing up to this limiting density. Instead, granular
systems of spheres arrest in a random close packing (RCP), which is
not optimal but occupies $\sim$64\% of space \cite{bernal}.  Previous
studies have derived geometric statistical models to map the
microscopic origin of the much debated 64\% density of RCP
\cite{bernal,torquato,ohern,liu,parisi,song,radin,clusel}. However,
the physical laws that govern its creation and render it the most
favorable state for randomly packed particles remains one of the most
salient questions in understanding all of jammed matter
\cite{torquato,ohern,liu,radin}. For instance, while it is known that
systems in equilibrium follow energy minimization and entropy
maximization to reach a steady state, the mechanism by which RCP is
achieved is much sought after.


Here we propose a thermodynamic view of the sphere packing problem
where the experimentally observed RCP can be viewed as a manifestation
of a singularity in a first-order phase transition.
Despite the inherent out-of-equilibrium nature of granular matter, the
formation of a jammed crystal can be mapped to a thermodynamic process
that occurs at a precise compactivity where the volume and entropy are
discontinuous.


We investigate mechanically stable packings ranging from the lowest
possible volume fraction of random loose packing (RLP) \cite{onoda} to
FCC. We numerically generate packings of $N=$10,000 spherical
particles of radius $R=100 \mu$m in a periodically repeated
cube. Initially, we use the Lubachevsky-Stillinger (LS) \cite{LS1,LS2}
and force-biased (FBA) \cite{FBA} algorithms to generate amorphous
configurations of unjammed hard-spheres fluids at infinite kinetic
pressure and volume fraction $\phi_i$ \cite{parisi,LS2} (see
Appendix-Section \ref{algorithm}).
While these configurations are geometrically jammed, they are not
jammed in a mechanical sense since the particles do not carry any
forces: the confining stress $\sigma$ (not kinetic) is zero. In fact
a key difference between granular materials jammed under external
stress or gravity and hard-sphere fluids is that, in the former,
each particle satisfies force and torque balance.
In order to study mechanically stable packings characterized by a
jamming transition we introduce interparticle forces via the
Hertz-Mindlin model of normal and tangential forces allowing the
particles to be soft but with a large Young modulus, $Y$. We then
simulate the process of jamming by Molecular Dynamics (MD) simulations
using previously developed methods \cite{makse} to compress the LS and
FBA packings from $\phi_i$ to a final jamming density,
$\phi_j$. Ultimately, we obtain mechanically stable packings just
above the jamming transition (in the limit of vanishingly small
confining stress, or equivalently in the hard-sphere limit,
$\sigma/Y\to 0^+$) covering a range of $\phi_j$ from $\phi_{\rm
  rlp}=0.55$ to crystallization at $\phi_{\rm fcc}=0.74$.

\begin{figure}
  \centering { \vbox {
      \resizebox{16cm}{!}{\includegraphics{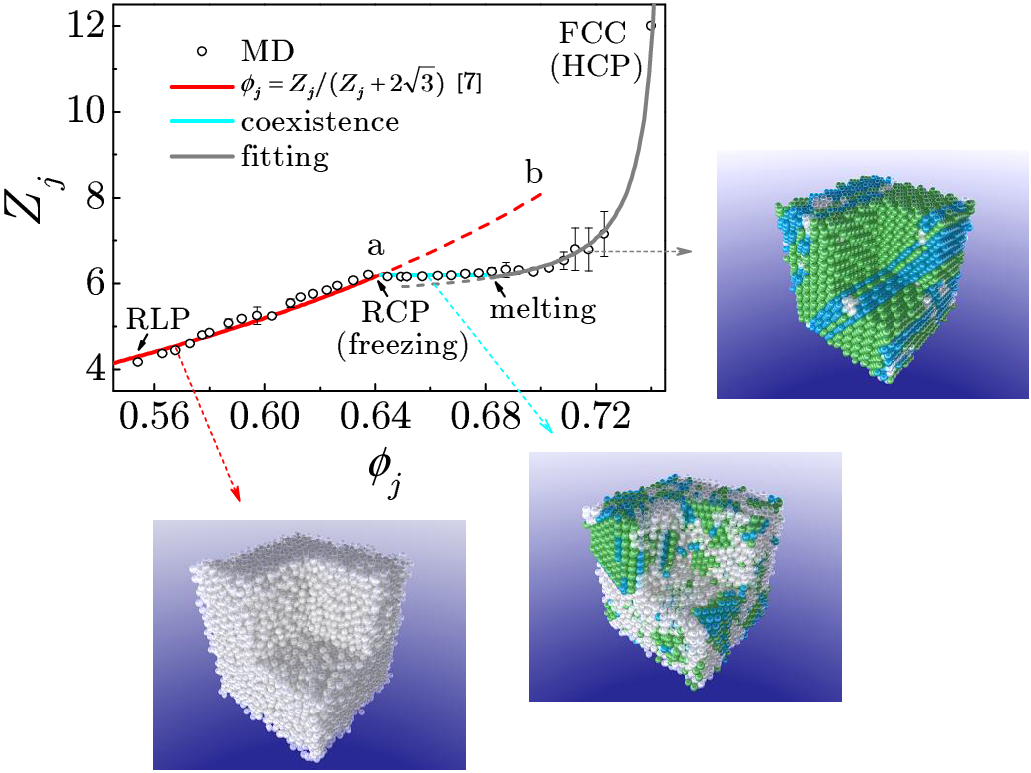}} } }
  \caption{}
\label{phase}
\end{figure}

The mechanical coordination number averaged over all the particles in
a packing, $Z_j$, characterizes different states of granular matter
\cite{bernal,aste,song}. Therefore, our study begins by plotting $Z_j$
versus $\phi_j$ for all generated packings.
Figure \ref{phase} suggests the existence of a transition occurring at
RCP evidenced by the abrupt plateau in $Z_j$. This transition could be
thought of as an analogue to the classical hard sphere liquid-solid
phase transition in thermal equilibrium
\cite{alder-wainwright1957,debenedetti}.  Such an analogy becomes
apparent if one identifies $Z_j$ of the jammed packing with the
kinematic pressure of the equilibrium hard sphere system \cite{parisi}
and it is in agreement with a recent conjecture regarding the
definition of RCP \cite{radin}.

Figure \ref{phase} identifies two branches and a coexistence region:
{\it (i)} An ordered branch of crystallized states with $\phi_j$
ranging from 0.68 to
a FCC lattice at 0.74. {\it (ii)} A disordered branch within
$0.55 \sim 0.64$ which can be fitted with the statistical theory
of \cite{song}: $\phi_j = Z_j/(Z_j +2 \sqrt{3})$ as shown in the
figure.
{\it (iii)} A coexistence region between 0.64 to 0.68 displaying a
plateau at the isostatic coordination number, $Z_{\rm iso}=6$
\cite{ohern,makse,moukarzel}. The intersection between the disordered
branch and the coexistence line identifies the ``freezing point'' of
the transition providing a definition of RCP.  Using the theoretical
results of \cite{song}, freezing occurs at $Z_{\rm iso}=6$ and
$\phi_{\rm rcp}=6/(6+2\sqrt{3})\approx 0.634$. The corresponding
``melting point'' appears at the other end of the coexistence at
$\phi_{\rm melt}=0.68$, signaling the beginning of the ordered
branch. Finite size analysis is shown in the Appendix-Fig. \ref{finite}A:
the results for 500 and 10,000 spheres are consistent with each other.
Other geometric aspects of the transition
are discussed in Appendix-Section \ref{other}.


To reveal the nature of the newly found phases we start with a
descriptive viewpoint and then turn to a thermodynamic analysis to
model the transition. In order to investigate if the concept of
phase transition applies to the trend observed in $Z_j$, one
commonly looks at the global ($Q_l, W_l$), and local ($q_6$)
orientational order parameters for a signature of varying amounts
of crystallization present in the packings as defined elsewhere
\cite{nelson} (see Appendix-Section \ref{order} and Fig.
\ref{descriptive} for definitions).
The salient feature of $Q_l$ is that its
zero value means disorder and non-zero value means
crystallization. Therefore, the presence in Fig. \ref{descriptive}A of
an increase in $Q_l$ from zero at $\phi_{\rm rcp}$ defines the
beginning of the coexistence region.
Typically, a first-order transition is marked by a nonzero third-order
invariant $W_l$ \cite{nelson,binder} which we find appears at the
melting point $\phi_{\rm melt}$ signaling the onset of the ordered
branch (Fig. \ref{descriptive}B).

Interestingly, the ordered phase has two significant peaks in the
probability distribution $P(q_{6})$ of the local order parameter
of each particle, $q_{6}$ (Fig. \ref{descriptive}C).  The peaks
correspond to FCC and HCP \cite{nelson} signaling that both
crystalline configurations are present in the ordered structure.
From the available data we cannot rule out the possibility of
another transition from HCP to FCC before $\phi_j\sim0.74$. The
Gaussian distributions $P(q_6)$ obtained for $\phi_j$ within 0.55
$\sim$ 0.64 show no preferred lattice structure in the random
branch.  While the relative peak positions in $P(q_6)$ do not
change, the percentage of crystal and random phase found in the
packing progresses from one to the other in the coexistence
region.  Microscopically, the existence of the two pure phases is
starkly represented by the two separated distributions of local
Voronoi volume fractions $P(\phi_{\rm vor})$ for which the same
phenomenology of $P(q_6)$ applies (Fig. \ref{descriptive}D,
$\phi_{\rm vor}=V_g/V_{\rm vor}$, where $V_{\rm vor}$ is the
Voronoi volume \cite{aste,song} of each particle of volume $V_g$).
This descriptive analysis is further supported in Appendix-Section
\ref{descriptive_picture}.

Having identified the structure of the phases we now develop a
thermodynamic viewpoint of the RCP transition to rationalize the
obtained results.
Transitions in equilibrium physical systems are driven by a
competition of energy and entropy.
Instead, a transition in athermal jammed matter is driven by the
minimization of the system's volume $W$ by compactification and
entropy maximization of jammed configurations
\cite{sirsam,ciamarra}. In accordance with the second law of
thermodynamics, the granular system tends to minimize its
Gibbs-Helmholtz ``free energy'' $F=W-X S$ rather than $W$ alone.
The compactivity of the system $X=dW/dS$ is a measure of how much
further compaction a packing can undergo; the lower the volume the
lower the compactivity \cite{sirsam}.  Thus,
we map the packing problem
to a thermodynamic problem where the volume $W$ replaces the energy
and $X$ takes the role of temperature. The principle of free energy
minimization can thus be applied.




The equations of state, $\phi_j(X)$ and $S(\phi_j)$, can be
calculated from the fluctuations of the Voronoi volumes in the
disordered and ordered phases \cite{chris}, $\sigma_{1}(\phi_j)$
and $\sigma_{2}(\phi_j)$ respectively, in analogy to the standard
Boltzmann statistical mechanics ($\sigma^2 \equiv \langle w^2_{\rm
  vor} \rangle - \langle w_{\rm vor} \rangle^2$ and $\omega_{\rm
  vor}=1/\phi_{\rm vor}=V_{\rm vor}/V_g$ is the reduced Voronoi
volume). Figure \ref{fluctuations}A shows clearly the existence of
the two pure phases and a discontinuity between both branches. We
obtain the compactivity by integration of $\sigma_{1}$ and
$\sigma_2$ using Einstein fluctuation theory
\cite{nowak,swinney,chris} (see Appendix-Section \ref{free} for more
details, we set $k_B=1$ for simplicity, $X$ is given in units of
$V_g$ and entropy is dimensionless):


\begin{subequations}
\begin{align}
  \frac{1}{X(\phi_j)}=\frac{1}{V_g}\int_{\phi_{\rm rlp}}^{\phi_j}
  \frac{d\phi}{\phi^2\sigma_{1}^{2}(\phi)}+\frac{1}{X_{\rm rlp}},
  \,\,\, \,\,\,\,\, \phi_{\rm rlp} \leq \phi_j \leq \phi_{\rm rcp}, \\
  \frac{1}{X(\phi_j)}=\frac{1}{V_g}\int_{\phi_{\rm
      melt}}^{\phi_j}\frac{d\phi}{\phi^2\sigma_{2}^{2}(\phi)}+\frac{1}{X_{\rm
      melt}}, \,\,\,\,\,\,\,\, \phi_{\rm melt} \leq \phi_j \leq
  \phi_{\rm fcc},
\end{align}
\label{x}
\end{subequations}
\noindent
where $X_{\rm melt}=X(\phi_{\rm melt})$ and $X_{\rm rlp}=X(\phi_{\rm
  rlp})$ are the compactivities at the melting point and at RLP,
respectively.
Once $X(\phi_j)$ is obtained from Eq. (\ref{x}), the entropy density,
$s=S/N$, is calculated by integration \cite{chris} (see Appendix-Section
\ref{free}):
\begin{subequations}
\begin{align}
 s(\phi_j)=s_{\rm rcp}+ V_g \int_{\phi_j}^{\phi_{\rm rcp}}
  \frac{d\phi}{X(\phi)\phi^{2}},
\,\,\,\,\,\,\,\, \phi_{\rm rlp} \leq \phi_j
  \leq \phi_{\rm rcp},
\\
  s(\phi_j)=V_g \int_{\phi_j}^{\phi_{\rm fcc}}
  \frac{d\phi}{X(\phi)\phi^{2}}, \,\,\,\,\,\,\,\, \phi_{\rm melt} \leq
  \phi_j \leq \phi_{\rm fcc},
\end{align}
\label{s}
\end{subequations}
where we have used that the entropy of FCC is zero in the
thermodynamic limit.  There are three unknown integration constants in
Eqs. (\ref{x}) and (\ref{s}): $X_{\rm rlp}$, $X_{\rm melt}$ and the
entropy of RCP, $s_{\rm rcp}$.  To close the system, we consider the
conditions for equilibrium between the phases \cite{sirsam}: (a)
``thermal'' equilibrium $X_{\rm melt}=X_{\rm rcp}\equiv X_c$, where $X_c$ is
the critical compactivity at the transition, and (b) the equality of
the free energy density (or chemical potential), $f=F/N$,
at the melting and the freezing RCP point: $f_{\rm melt}=f_{\rm
  rcp}$. This implies, $\omega_{\rm rcp} - (X_c/V_g) s_{\rm rcp} =
\omega_{\rm melt} - (X_c/V_g) s_{\rm melt}$, where
$w=1/\phi_j=W/(NV_g)$ is the reduced volume of the system.  The third
integration constant $X_{\rm rlp}$ can be considered infinite
\cite{song,swinney,chris} since RLP is the highest volume of the
system. While the precise value of $X_{\rm rlp}$ does not affect our
conclusions, a more accurate finite value can be obtained by fitting
the entropy Eq. (\ref{s}) with an independent measure obtained by
cluster analysis from information theory (Shannon entropy, $s_{\rm
  shan}$) as developed in \cite{chris} (Appendix-Section \ref{en}).
Figure \ref{fluctuations}B shows that the entropy from the
thermodynamic integration Eq. (\ref{s}) and $s_{\rm shan}$ agree
well (up to a multiplicative constant)
supporting the framework of Eqs. (\ref{x})-(\ref{s}).  The entropy is
composed of two branches plus the coexistence region (green line in
Fig. \ref{fluctuations}B).


Figure \ref{compactivity}A displays a discontinuity in $s(X)$ at
$X_c=0.031 V_g$ revealing the first-order nature of the transition
which is accompanied by an ``entropy of fusion'' $\Delta s_{\rm
fus} \equiv s_{\rm rcp} - s_{\rm melt}=3.0$.  The volume fraction
is discontinuous at $X_c$ (Fig. \ref{compactivity}B) where the
system jumps from RCP to the melting point
releasing an amount of volume given by the ``enthalpy of fusion''
$\Delta h_{\rm fus} = X_c \Delta s_{\rm fus} = 0.09 V_g$ while the
compactivity stays constant.  This process corresponds to the
typical latent heat in exothermic first-order transitions.

Systems jammed at RCP need to overcome a volume barrier $\Delta h_{\rm
  fus}$ for crystal formation or, equivalently, particle displacements
$\Delta r_{\rm fus} \approx 0.45 R$.
From a thermodynamic perspective, the requirement is equivalent to
bringing a random packing in contact with a compactivity bath at
$X<X_c=0.031 V_g$. The fundamental idea is to surround a random
packing above $X_c$ with a crystal lattice below $X_c$ and perturb
the system to equilibrate.  A shear cycling experiment---which
conserves the shape of the box containing the particles---suffices
to explore the crystal branch
\cite{scott-crystal,pouliquen-shear}. Shear-induced
crystallization has been observed
\cite{scott-crystal,pouliquen-shear}
when the maximum angle of horizontal shear is above $\theta\approx
10^{\circ}$. This value is of the same order as our estimate of
the shear amplitude to crystallize at $\phi_j=68\%$ based on the
entropy of fusion, which gives $\theta \equiv \tan^{-1}(\Delta
r_{\rm fus}/2R) \approx 13 ^{\circ}$. Furthermore, recent shear
cycling experiments \cite{frank2,radin} appear to be in reasonable
agreement with the present results.
We also expect that 2d equal-sized disks may have a near zero
entropy of fusion owning to their tendency to easily crystallize
while $\Delta s_{\rm fus}$ may sharply increase in 4d and above
\cite{parisi}.

The behavior of the free energy density shown in Figs.
\ref{compactivity}C and \ref{compactivity}D summarizes the mechanism
to achieve RCP. The free energy in Fig. \ref{compactivity}C increases
as $X$ decreases from RLP to freezing at RCP. At $X_c$, the system
transitions to the phase with the lower free energy through an entropy
discontinuity given by $s=-\partial f/\partial X$. The system may also
enter the metastable branch as indicated in Fig. \ref{compactivity}C
and in Figs. \ref{phase} and \ref{fluctuations}B from $a \to b$.  In
the spirit of Landau mean field theory of phase transitions, we relate
the distribution of the local order parameter to the free energy
functional, $\cal F$, and $X$ as $P(q_6)\approx \exp[-{\cal
  F}(q_6)/X]$ \cite{binder}.  Figure \ref{compactivity}D shows ${\cal
  F}(q_6)$ displaying the minima of ${\cal F}(q_6)$ defining the order
and disorder phases at different $\phi_j$.  We find that the location
of the minimum at $q^{\rm min}_6\approx 0.425$ remains constant from
RLP up to the freezing point as expected in the disordered phase. The
value of ${\cal F}(q^{\rm min}_6)$ is very deep for $\phi_j=0.55$ and
becomes less deep as the freezing point is approached. The value of
${\cal F}(q^{\rm min}_6)$ at the freezing and melting points become
approximately equally deep indicating the phase coexistence at $X_c$.
Within a statistical mechanics framework, these results are a natural
consequence and give support to such an underlying statistical
picture.

In conclusion, treating granular packings from the perspective of
theoretical physics developed by Boltzmann and Gibbs has the potential
to answer basic questions in the field of disordered media. State
variables like the compactivity can be introduced with the potential
of identifying transition points between different phases, a fact that
can unequivocally define RCP as the freezing point in a discontinuous
transition.
This formalism may be useful in analyzing other related
transitions in complex systems ranging from optimization problems
in computer science \cite{kurchan} to the physics of glasses
\cite{parisi}. Other unsolved packing problems including finding
the densest arrangement of rods, ellipsoids, spherocylinders,
binary mixtures, Platonic and Archimedean solids---which are known
to pack better than spheres \cite{glotzer,clusel}---can now be
analyzed from the proposed thermodynamic view of phase
transitions.



\vspace{0.5cm}

{\bf Acknowledgements}. This work is supported by the National Science
Foundation. We are grateful to B. Bruji\'c, M. Shattuck, F. Zamponi,
C. Song and P. Wang for discussions.

\clearpage

FIG. \ref{phase}. The RCP transition. We plot the mechanical
coordination number $Z_j$ versus the volume fraction $\phi_j$ for
each packing. We identify: {\it (i)} a disordered branch which can
be fitted with the statistical model of \cite{song} as shown, {\it
(ii)} a coexistence region, and {\it (iii)} an ordered branch.
Error bars are calculated over 523 packings obtained from initial
LS configurations. The 3d plots visualize how the transition
occurs in terms of arrangements of contacting particles. White
particles are random clusters, light blue are HCP and green are
FCC clusters. Further microscopic information regarding the
transition is provided in Appendix-Section \ref{geometry}.

FIG. \ref{descriptive}. Descriptive viewpoint of the RCP
transition. {\bf (A)} Global orientational order parameters $Q_l$
versus $\phi_j$ for different packings signaling the freezing
point at $\phi_{\rm rcp}$. A linear fit is possible in the
coexistence region. {\bf (B)} Global third-order invariants
$W_{l}$ versus $\phi_j$ signaling the melting point at $\phi_{\rm
melt}$. {\bf (C)} Probability distribution of local orientational
order parameter $P(q_{6})$ versus $q_6$ (vertical axis) for
packings with $\phi_j$ (horizontal axis).  For packings with
$\phi_j$ within 0.68 $\sim$ 0.72, the distributions have two
significant peaks centered at $q^{\rm
  fcc}_{6}=0.57$ and $q^{\rm hcp}_{6}=0.48$, which correspond to FCC
and HCP arrangements, respectively \cite{nelson}.  Color bar
indicates the values of $P(q_{6})$. Since the peaks are very
pronounced, we plot $P(q_{6})$ up to the indicated value.  {\bf
(D)} Probability distribution of local volume fractions of the
Voronoi volumes of each particle, $P(\phi_{\rm vor})$ versus
$\phi_{\rm vor}$ (vertical axis) for different $\phi_j$
(horizontal axis). The plot indicates a clear discontinuity
between both branches. Color bar indicates the values of
$P(\phi_{\rm vor})$ which are plotted up to the indicated value.


FIG. \ref{fluctuations}.  Equations of state of the RCP
transition. {\bf (A)} Volume fluctuations of the Voronoi cell of a
particle as a function of $\phi_j$.  The data indicates a
discontinuity between the ordered and disordered branches which
are fitted by functions as indicated. These fittings are used in
the integrations of Eq. (\ref{x}). The larger fluctuations in
volume observed in the order state compared to the disorder state
at similar $\phi_j$ are due to the fact that the system packs
better in the former and thus displays larger fluctuations when
the system volume is the same. {\bf (B)} Entropy obtained from
fluctuation theory in Eq. (\ref{s}), $s$, and Shannon entropy from
information theory, $s_{\rm shan}$, versus volume fraction
$\phi_j$. Both entropies agree (up to a multiplicative constant,
$k=0.1$, as indicated) confirming our calculations. The extended
branch denotes a metastable state ending at point $b$ at an
hypothetical Kauzmann density, $\phi_K$, in analogy with the
physics of glasses \cite{parisi} (see Appendix-Section \ref{glass}).

FIG. \ref{compactivity}. Thermodynamic viewpoint of the RCP
transition. All the observables are consistent with a transition
at $X_c=0.031 V_g$. {\bf (A)} Entropy $s$ versus $X$.  {\bf (B)}
Volume fraction $\phi_j$ versus $X$.  {\bf (C)} Free energy
density $f$ versus $X$.  We extend $f$ for both branches to
indicate the possible metastable states. At $X_c$ the system
follows the minimization of the free energy signaling the
transition from RCP to order.  {\bf (D)} Free energy functional
${\cal F}(q_6)$ versus $q_6$ (vertical axis) $\phi_j$ (horizontal
axis). Color bar indicates the values of ${\cal
  F}(q_6)$, which are plotted in the range indicated to focus on the
region of coexistence.  The minima correspond to the disordered phase
and the FCC and HCP phases in the ordered branch.

\clearpage

\clearpage

\begin{figure}
  \hbox{ {\bf (A)} \resizebox{8cm}{!}{\includegraphics{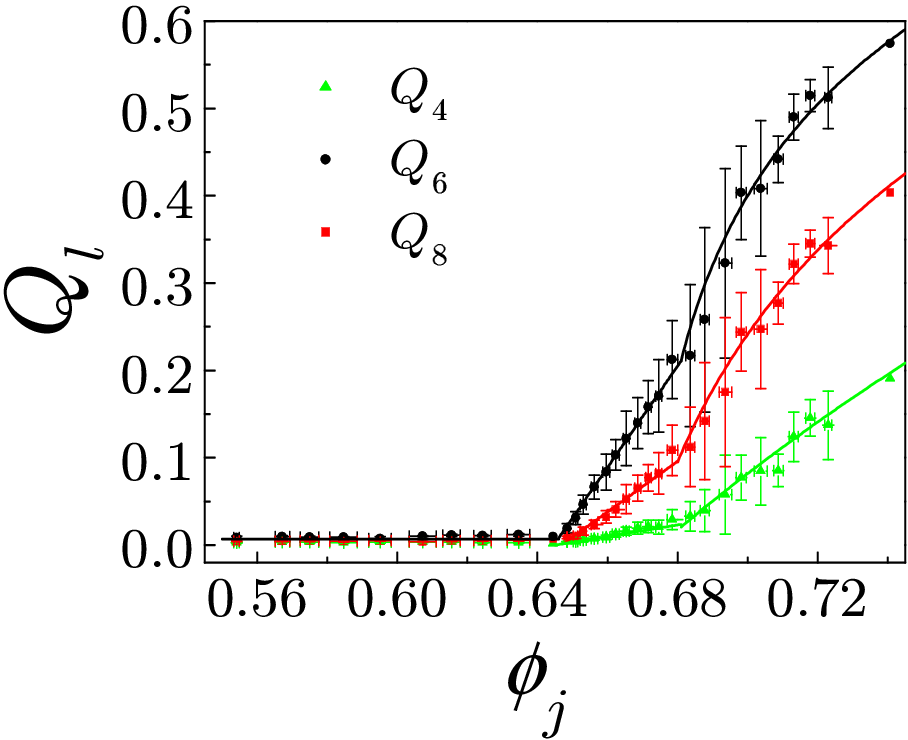}}
  { \hbox {\bf (B)}
      \resizebox{7.5cm}{!}{\includegraphics{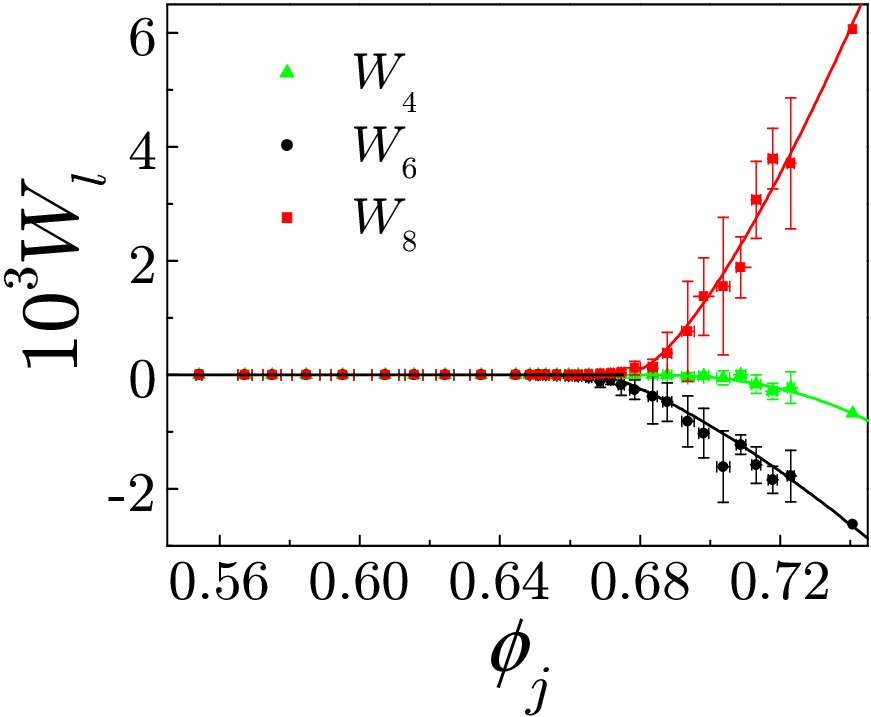}} }}
  \vspace{.5cm}    
\vbox{ {\bf
      (C)} \resizebox{15cm}{!}{\includegraphics{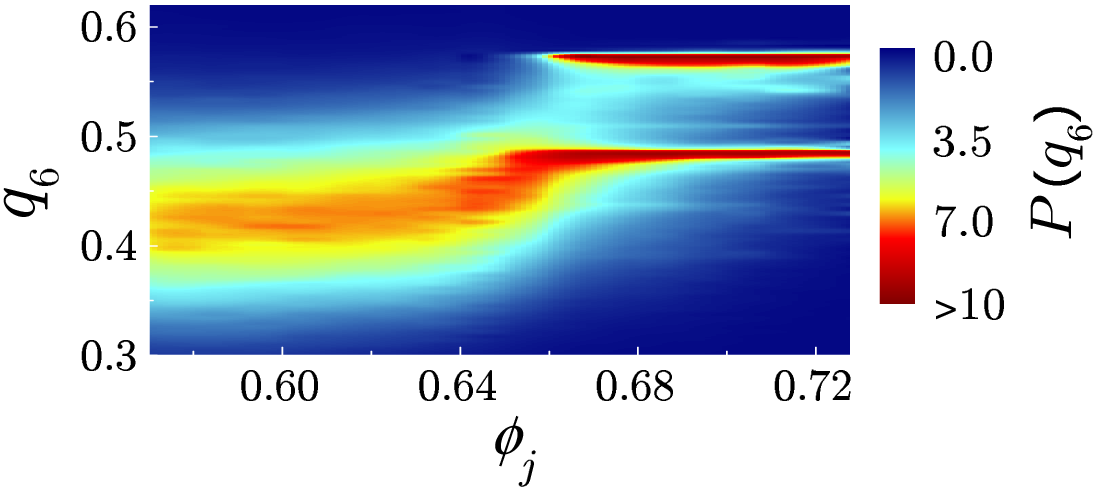}}
  } \vbox{ {\bf (D)}
    \resizebox{15cm}{!}{\includegraphics{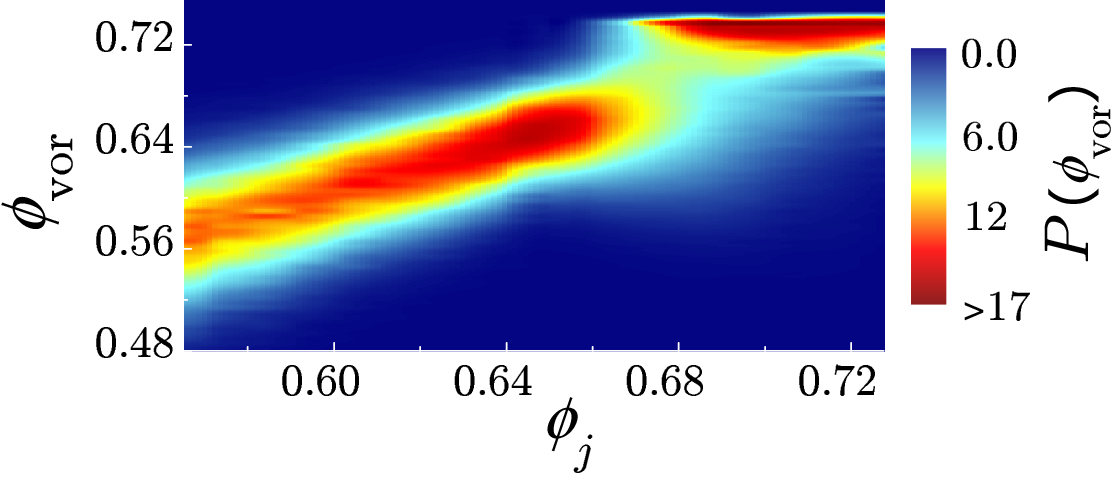}} }
  \caption{}
\label{descriptive}
\end{figure}

\clearpage

\begin{figure}
\vbox{  {\bf (A)} \centering \resizebox{11cm}{!} {\includegraphics{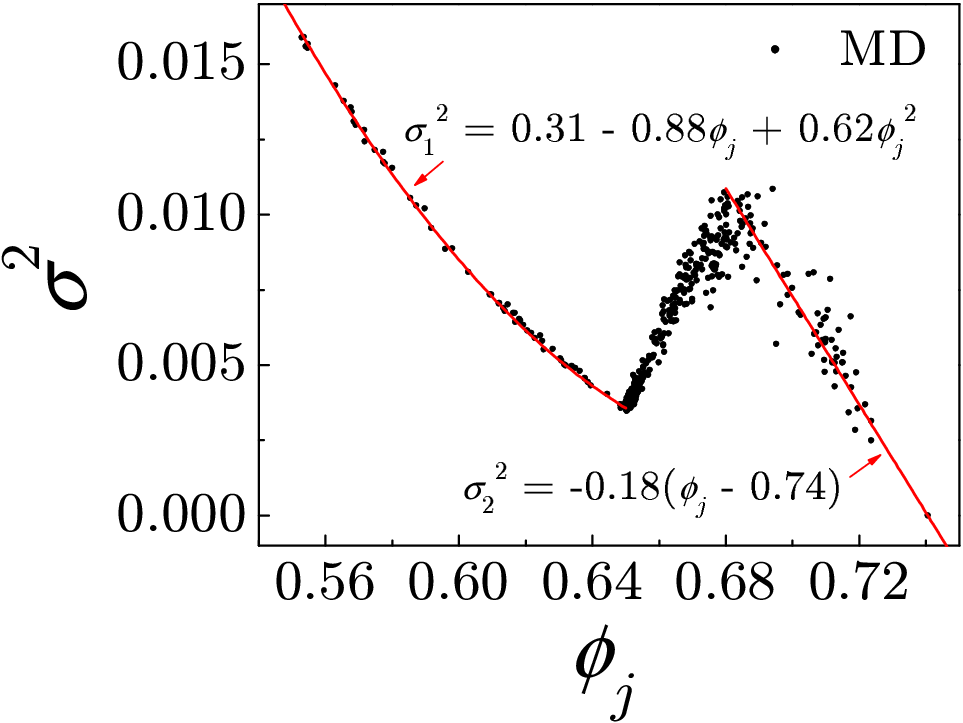}}}
\vbox{ {\bf (B)}
  \centering \resizebox{10cm}{!} {\includegraphics{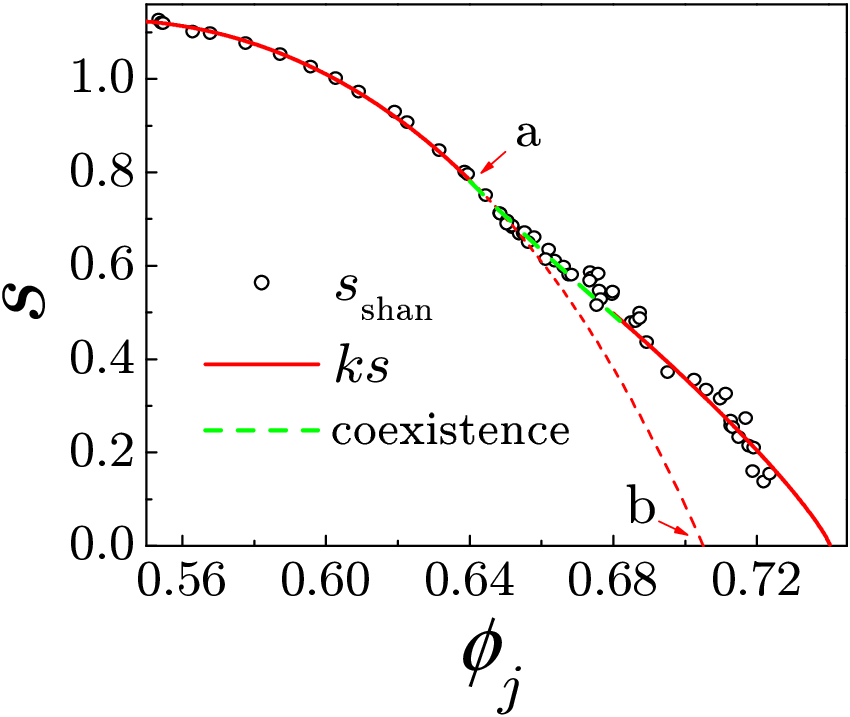}}}
\caption{} \label{fluctuations}
\end{figure}

\clearpage

\begin{figure}
  \hbox{ \centering {\bf (A)} \resizebox{7.5cm}{!}
    {\includegraphics{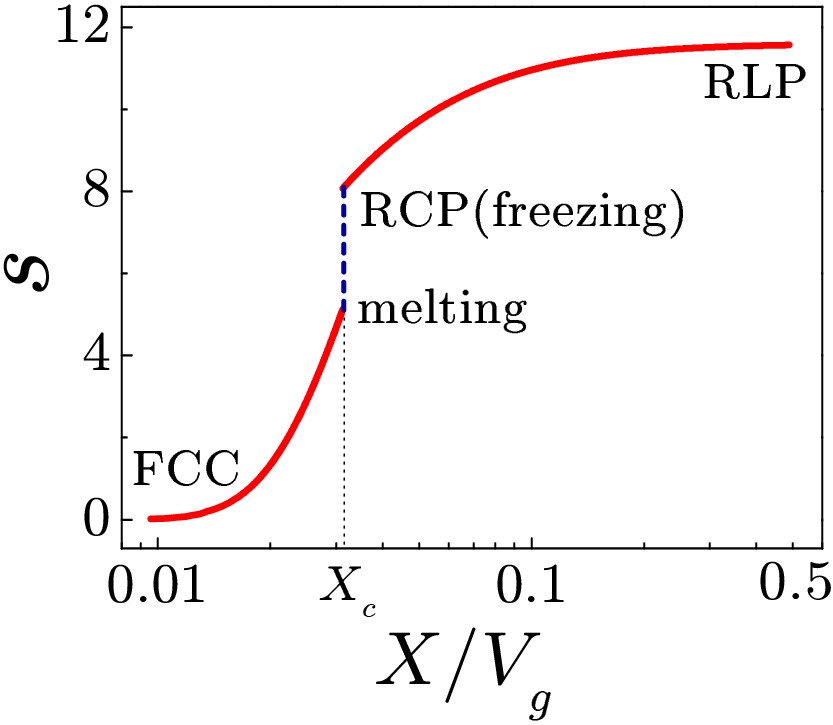}} \centering {\bf (B)}
    \resizebox{8.5cm}{!}  {\includegraphics{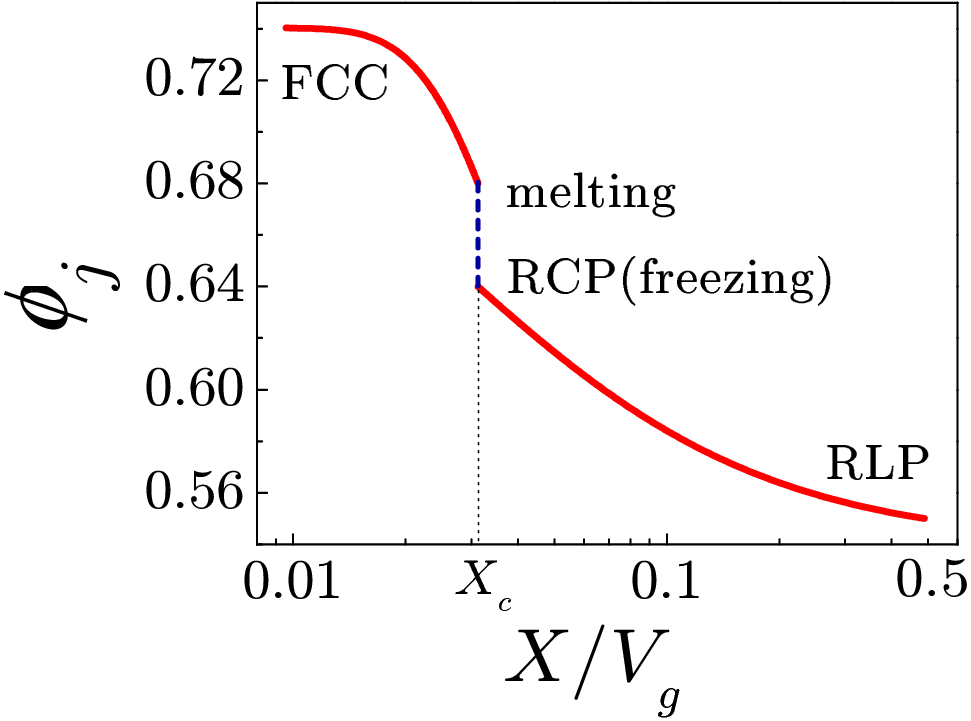}} }
  \vspace{.5cm}
\centerline {  \hbox{ \centering {\bf (C)} \resizebox{9cm}{!}
    {\includegraphics{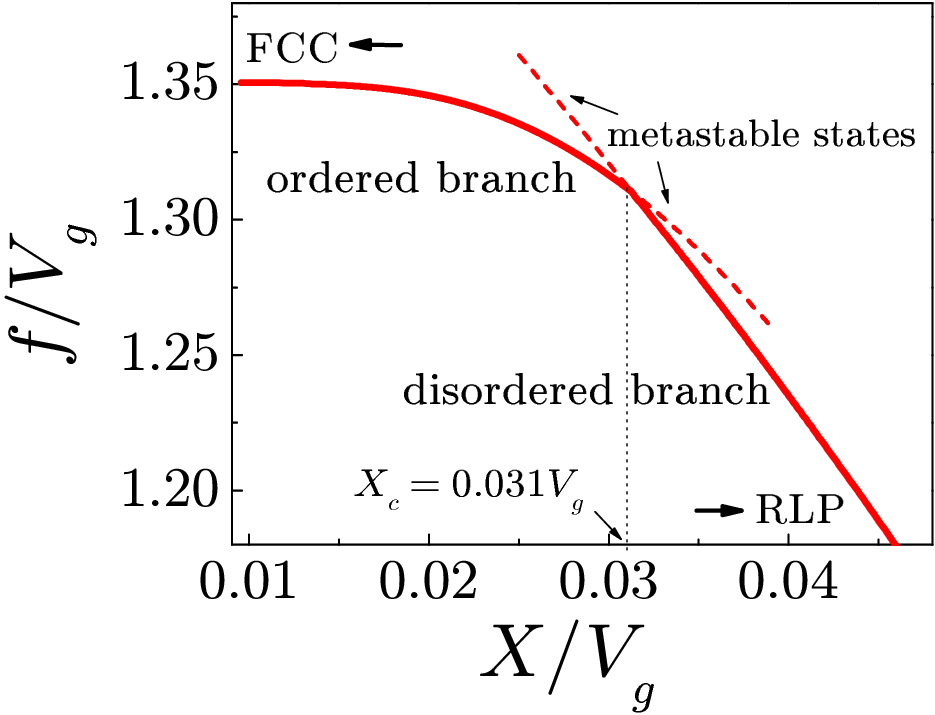}} } }
\hbox{
\centering {\bf (D)} \resizebox{16cm}{!}
    {\includegraphics{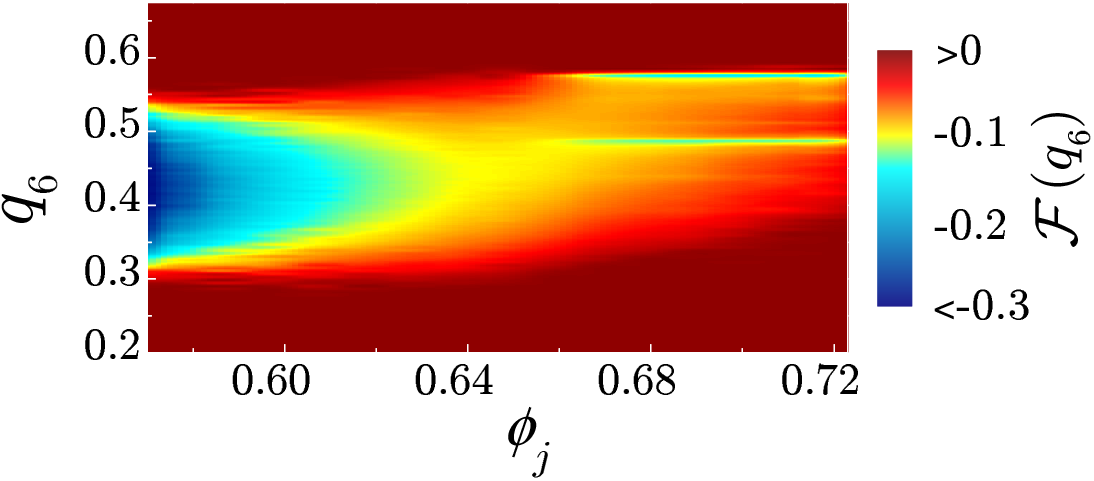}}}
  \caption{}
  \label{compactivity}
\end{figure}

\clearpage

\centerline{\bf Appendix}

\vspace{.3cm} { \bf A first order phase transition at the
  random close packing of hard spheres}

\vspace{.3cm} \centerline{Yuliang Jin and Hern\'an A. Makse}

\vspace{.4cm}

Here, we describe the details of the MD simulations (Section
\ref{algorithm}), geometrical interpretations of the transition
(Section \ref{other}), and the calculations leading to the descriptive
(Section \ref{descriptive_picture}) and the thermodynamic (Section
\ref{thermo}) view of the RCP transition.

\section{Algorithm}
\label{algorithm}

We use computer simulations to obtain jammed packings containing
$N=10,000$ mono-disperse spheres of radius $R=100 \mu$m with
periodic boundary conditions.  We first apply a modified
Lubachevsky-Stillinger (LS) algorithm \cite{LS1,LS2} to generate
packings of densities up to $\sim$0.72. In the algorithm, a set of
random distributed points grow into nonoverlaping spheres at a
fixed expansion rate $\gamma$. The spheres are considered
perfectly elastic and evolve in time according to Newtonian
dynamics. The configurations eventually arrive at
out-of-equilibrium states with a diverging collision rate and a
density $\phi_i$.  Practically, we set the reduced kinetic
pressure of the fluid defined as $p=PV/Nk_BT$ to be $10^{12}$
\cite{LS2} as a criteria of the diverging collision rate. The
final packing configurations depend on the expansion rate
$\gamma$: large values of expansion rate result in random packings
with very low packing densities, while small values of expansion
rate result in packings with higher densities.

Although the packings obtained from the modified LS algorithm are
considered as ``geometrically jammed", they are not jammed in the
mechanical sense since the particles do not carry on any forces.
In order to study mechanical stable packings characterized by a
jamming transition, we model the microscopic interaction between
deformable grains by the nonlinear Hertz-Mindlin normal and
tangential forces \cite{makse,song}. We use configurations from
the modified LS as the starting point, $\phi_i$, and apply
molecular dynamics to simulate Newton equations for the evolution
of the particles following algorithms in \cite{makse,song}.

The aim of this part of the protocol is to generate mechanically
stable jammed packings at the jamming transition $\phi_j$.  For the
packings obtained by the LS algorithm, we first reset the velocities
of the particles to zero. At this point the system stress $\sigma$ and
mechanical coordination number $Z_j$ are zero since there is no
deformation or overlapping between the particles.  We notice that the
stress $\sigma$ is not the kinetic pressure, $p$, measured in the LS
packings which diverges at the end of the LS protocol. Here $\sigma$
refers to the mechanical pressure related to the trace of the stress
tensor $\sigma_{ij}$ via $\sigma = \sigma_{ii}/3$, where

\begin{equation}
  \sigma_{ij} = \frac{R}{2V} \sum_{\rm contacts} f^c_{i} n^c_{j} + f^c_{j} n^c_{i},
\end{equation}
where the sum is over all the contact forces, $f^c_{i}$ denotes the
$i$-th component of the contact force,  $\hat {\bf n}^c$ is the unit
vector joining the center of two spheres of radius $R$ in contact and
$V$ is the system volume.

The system is then compressed isotropically by a constant compression
rate until a given nonzero stress $\sigma$ is reached. Next, we turn
off the compression and allow the system to relax with constant
volume. If the system eventually reaches a jammed state with a fixed
nonzero $\sigma$ and coordination number, the system pressure will
remain unchanged over a large period of time (usually $\sim10^7$ MD
steps); otherwise, if the system is not stable, the pressure will
relax to zero very fast \cite{song}.

Previous studies show that there exists a jamming transition for
granular matter as
\begin{equation}
\sigma(\phi)-\sigma_j\sim(\phi-\phi_j)^\alpha,
\label{sig}
\end{equation}
and
\begin{equation}
  Z(\phi)-Z_j\sim(\phi-\phi_j)^\beta.
\end{equation}
Here, $\sigma_j$ is zero for frictionless packings. However, it
could has nonzero value for frictional packings, in general. The
exponent $\alpha=3/2$ is trivially related to the Hertz-law of
interparticle contact force and $\beta=1/2$ seems to be universal
over different force laws \cite{ohern}.
\begin{figure} \centering { \hbox { {\bf (A)}
\resizebox{7.5cm}{!}{\includegraphics{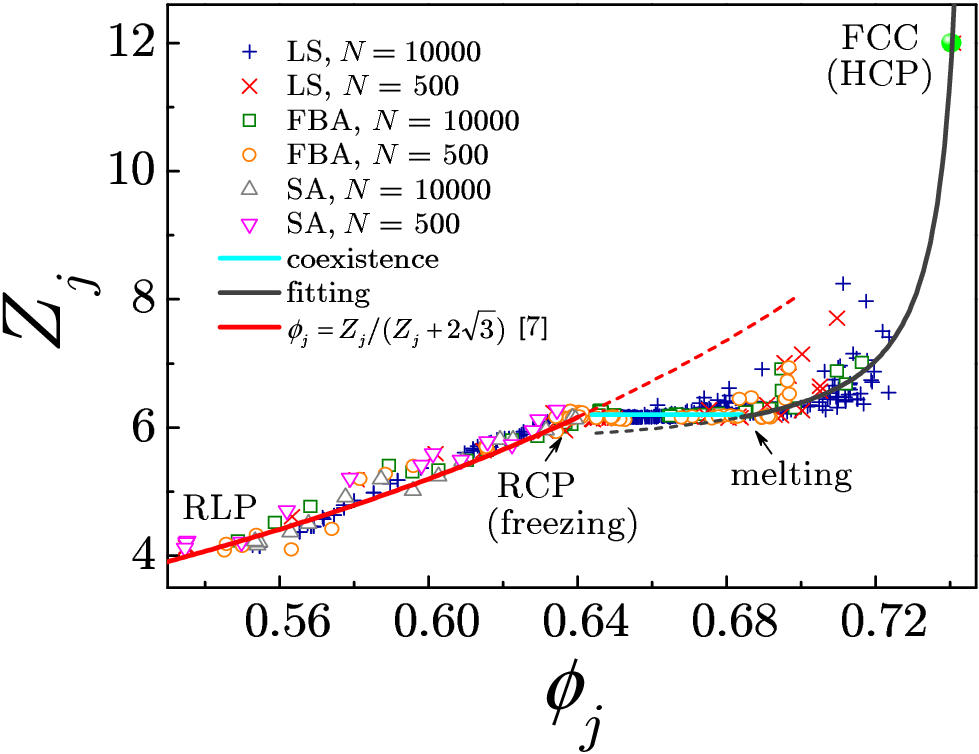}} {\bf (B)}
\resizebox{7cm}{!}{\includegraphics{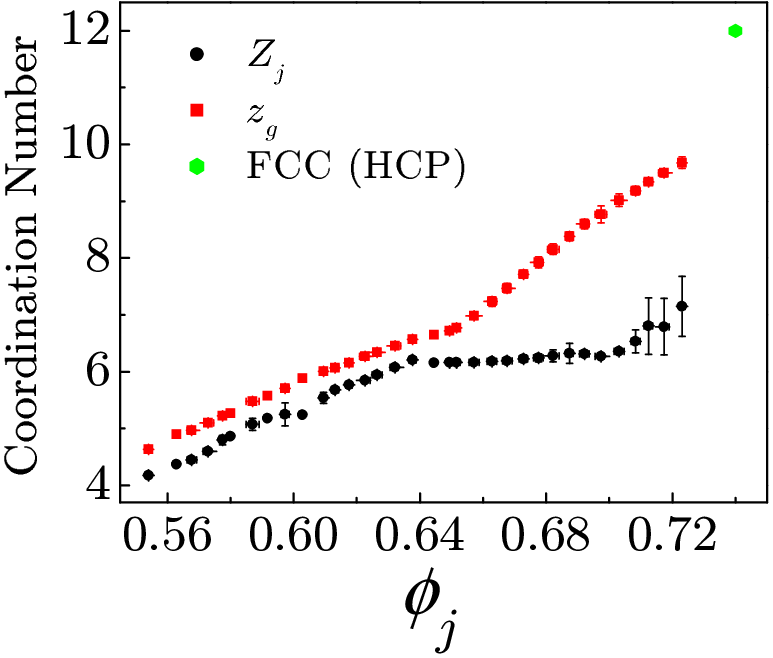}} } } \caption{
{\bf (A)} The mechanical coordination number $Z_j$ versus the
volume fraction $\phi_j$ with different system sizes (500 and
10,000 particles) and algorithms (Lubachevsky-Stillinger LS,
force-biased FBA, and split SA algorithms). The results show that
the transition does not depend on the system size and the
algorithm used. {\bf (B)} Comparison between geometrical
coordination number $z_g$ and mechanical coordination number
$Z_j$. Along the disordered branch, $z_g$ closely follows the
mechanical coordination number $Z_j$. We expect that in the
thermodynamical limit the gap between both coordinations may
diminish. $z_g$ and $Z_j$ start to diverge at RCP, which is an
indication of increasing geometric degeneracies in the contact
network at the onset of crystallization.} \label{finite}
\end{figure}

In practice, it is difficult to reach a jammed packing exactly at
the transition point $\phi_j$ while it is much easier to get a
stable packing with slightly higher pressure. In order to approach
the transition point, a jammed packing at higher pressure than
$\sigma_j$ in Eq. (\ref{sig}) obtained using the above protocol is
decompressed with a negative compression rate until certain lower
pressure is reached. Then the system is allowed to relax again to
check for mechanical stability. If it is stable, then the system
is decompressed further to an even lower pressure, and we check
its stability again. By this process (called the split algorithm
in \cite{song}) we are able to approach the density $\phi_j$ at
the jamming transition point as close as possible within the
system error. The pressure of the packings at the jamming point
studied in this paper is $100\pm8$ KPa. The difference between the
volume fraction $\phi$ of these packings and the critical volume
fraction $\phi_j$ from power-law fitting in Eq. (\ref{sig}) is
about $10^{-3}$.

The same preparation protocol is repeated by using the force-biased algorithm (FBA) of \cite{FBA} as initial protocol. The force-biased algorithm is a variant of the original method of W. S. Jodrey and E. M. Tory, {\it Phys. Rev. A} {\bf 32}, 2347 (1985),  We also generate packings following the split algorithm of \cite{song} starting with low initial volume fractions at $\phi_i=0.3$ below RLP.

The mechanical coordination number, $Z_j$, versus the final
jamming density, $\phi_j$, is plotted in Fig. \ref{finite}A for
all the obtained jammed packings which total 720.  The plot
signals the existence of a transition at RCP.  It is analogous to
the equilibrium liquid-solid transition in hard-spheres, if we
replace $Z_j$ by the kinetic pressure of the fluid \cite{parisi}.

\section{Other aspects of the RCP transition}
\label{other}

\subsection{Finite size analysis}

It is important to determine the finite size effects of our
results. Figure \ref{finite}A shows the results for smaller
systems of 500 particles compared with 10,000 spheres system used
in Fig. \ref{phase}. We find that both plots are consistent with
each other.  While Fig. \ref{phase} shows the average of $Z_j$
over the LS packings, Fig. \ref{finite}A shows each point
representing a single packing obtained with the indicated
algorithms for different system sizes. We find that the transition
is similar over the different protocols.


\subsection{Isostatic and geometrical coordination numbers and
  symmetry breaking}

\label{geometry}

While the isostatic coordination number has been well documented at
RCP \cite{ohern,makse,moukarzel}, the possibility of states with
$Z_j=6$ along the coexistence region with $\phi_{\rm rcp} < \phi_j <
\phi_{\rm melt}$ requires more elaboration \cite{parisi}.  We recall,
however that the limiting condition $Z_{\rm iso}=6$ is necessary, but
not sufficient, for a rigid isostatic aggregate: a kinematic condition
for rigidity must hold \cite{moukarzel} where the functions that
describe the connections between the centers of contacting particles
are independent.  The presence of crystal-like regions suggests that
this condition may not be satisfied. Thus, the packings with $Z_j=6$ in
the coexistence region are not necessarily isostatic, except exactly at the
freezing point.

It is interesting to understand the geometrical rearrangements
occurring during the RCP transition in light of the fact that the
packings enter the coexistence region from RCP by keeping $Z_j=6$
constant.  The particles modify the positions of the 6 contacts in
average at RCP to create crystal-like regions without creating new
contacts or destroying old ones.  This implies that the
arrangements of particles are such that particles in the second
coordination shell come closer to the central particle and contribute
to the first coordination shell, yet without producing a new
contacting force since $Z_j$ is kept at 6 in the entire
coexistence region.  The new particles moving into the first
coordination shell can be considered in geometrical contact but
carrying no force. Thus following \cite{song} we introduce the
idea of geometrical contact, $z_g$, as those particles in the
first coordination shell that do not provide any force but still
are close enough to the central particle to contribute to the
geometrical contact network. The geometrical coordination number $z_g$ is different
from the mechanical coordination number $Z_j$ which only counts those contacts with
nonzero forces. By definition $z_g\geq Z_j$.

While $Z_j$ is easy to measure in computer simulations of soft
particles as the number of contacts between overlapping particles,
the geometrical coordination number, $z_g$, can be measured by
slightly inflating the spheres up to 4\% of their diameters and
counting the resulting contact particles, as discussed in
\cite{song}.  In practice, the geometrical coordination number
measures the particles surrounding a central one with a gap
between them from zero or negative (giving $Z_j)$ up to
$\delta=0.08R$ as discussed in \cite{song}. We notice that
$\delta=0.08R$ is much smaller than the location of the second
peak in the radial distribution function which occurs around
$\delta\sim 2R$. The value $\delta=0.08R$ is specific for a system
of $N$=10,000. We expect this value to diminish in the
thermodynamic limit.

Figure \ref{finite}B plots the geometrical and mechanical coordination
$z_g$ and $Z_j$ as a function of $\phi_j$ for the same packings as in
Fig. \ref{phase}. We find that along the disordered branch,
$z_g\approx Z_j$ as expected \cite{song}. However, in the coexistence
region, $Z_j= 6$ stays constant while $z_g$ keeps growing with
$\phi_j$. The separation between $z_g$ and $Z_j$ is a signature of the
onset of ordering at the freezing point. As explained above, the
system starts to crystallize by allowing particles in the second
coordination shell to come closer to the central particle and moving
the $Z_j$ contacting particles towards a FCC arrangement. At the
melting point, the condition $Z_j=6$ cannot hold any longer and the
system transitions to the other branch with an increase of $Z_j$ up to
12.

The distinction between $z_g$ and $Z_j$ is not only important for
a characterization of the transition. It is also important to
interpret the experimental results. Due to the uncertainty in
detecting the exact position of the particles in any experiment,
the exact mechanical coordination might be very difficult to
obtain. Thus, a small uncertainty in the determination of the
contacting particles $\delta=0.08R$ will produce $z_g$ as shown in
Fig. \ref{finite}B. One way to obtain the actual mechanical
coordination from experimental data is to use the experimentally
obtained coordinates of the particles as initial positions of a MD
simulation using Hertz-Mindlin forces to relax the configurations
and find the exact force balance network of the packing. Codes to
develop this procedure are available at www.jamlab.org. We also
provide most of the packings used in this study as well as the code to
calculate the entropy.

\subsection{Relation to the glass transition}
\label{glass}

The thermodynamic character of the RCP transition seemingly
contrasts to the non-thermodynamic viewpoint of the glass
transition which proposes a dynamical arrest upon supercooling
\cite{debenedetti}. However, the same phenomenology of
vitrification could be applied to the RCP transition by
extrapolating the entropy of the disordered branch $s(\phi_j)$
into a metastable region below the freezing point as schematically
shown in Figs. \ref{phase} and \ref{fluctuations}B from $a \to b$.
Two scenarios may emerge: the metastable branch may end in a
metastability limit at the spinodal $\partial X/\partial S=0$
\cite{debenedetti} or it may continue until the entropy of the
metastable liquid is zero as shown in Fig. \ref{fluctuations}B.
Such a scenario
would predict a Kauzmann density $\phi_K$ at which the entropy of the
disordered branch vanishes, signaling the existence of an ideal jammed
glass analogous to the Kauzmann temperature in glasses \cite{parisi}.

\section{Descriptive viewpoint of the RCP transition}
\label{descriptive_picture}

\subsection{Orientational order parameter}
\label{order}

The orientational order is measured by associating a set of
spherical harmonics with every bond joining a sphere and its
neighbors \cite{nelson}:

\begin{equation}
Q_{lm}\left(\vec{r}\right)=
Y_{lm}\left(\theta\left(\vec{r}\right),
\phi\left(\vec{r}\right)\right),
\end{equation}
where the $\{Y_{lm}(\theta,\phi)\}$ are spherical harmonics, and
$\theta(\vec{r})$ and $\phi(\vec{r})$ are the polar angles of the
bond.  The local orientational order parameter $q_{l}$ for a
particle $i$ is given by rotationally invariant combinations of
$Q_{lm}$,

\begin{equation}
q_{l}=\left[\frac{4\pi}{2l+1}\sum_{m=-l}^{l}\left|\overline{Q}_{lm,i}\right|^2\right]^{1/2},
\end{equation}
where $\overline{Q}_{lm,i}$ is averaged over $N_{i}$ neighbors of
this particle,

\begin{equation}
\overline{Q}_{lm,i}=\frac{1}{N_{i}}\sum_{j=1}^{N_{i}}Q_{lm}\left(\vec{r}_{ij}
\right).
\end{equation}

We also consider the global orientational order $Q_{l}$, as well as
the third-order invariants $W_{l}$, which are

\begin{equation}
Q_{l}=\left[\frac{4\pi}{2l+1}\sum_{m=-l}^{l}\left|\overline{Q}_{lm}\right|^2\right]^{1/2},
\end{equation}
where the average is taken over all the $N_b$ bonds in the packing,
\begin{equation}
\overline{Q}_{lm}=\frac{1}{N_{b}}\sum_{bonds}Q_{lm}\left(\vec{r}\right),
\end{equation}

and
\begin{equation}
W_{l}=\sum_{\substack{m_{1},m_{2},m_{3}\\m_{1}+m_{2}+m_{3}=0}}
\begin{bmatrix}
l&l&l\\m_{1}&m_{2}&m_{3}
\end{bmatrix}
\times
\overline{Q}_{lm_{1}}\overline{Q}_{lm_{2}}\overline{Q}_{lm_{3}},
\end{equation}
where the coefficients $\begin{bmatrix} l&l&l\\m_{1}&m_{2}&m_{3}
\end{bmatrix}$
are the Wigner $3j$ symbols.

We use a definition of bond as in \cite{nelson} where all spheres within $r_c=1.2d$ of a given sphere are near neighbors, where $d=2R$ is the sphere diameter. We note that $r_c=1.2d$ is the center of the first peak (centered at $r=0$) and the second peak (centered at $r=1.4d$) in the radial distribution function $g(r)$ of a perfect FCC lattice. Thus, this criteria is also consistent with the definition used in T. M. Truskett, S. Torquato, P. G. Debenedetti, {\it Phys. Rev. E} {\bf 62}, 993 (2000), where $r_c$ is the first minimum in $g(r)$. The neighbors can also be defined as those who have mechanical contacts with the given sphere, $Z_j$. The basic results do not change by changing the definition of nearest neighbors. The results of the global orientational order $Q_l$ and $W_l$ are shown in Fig.  \ref{descriptive}A and Fig. \ref{descriptive}B.  To investigate the structures of the phases, Fig. \ref{descriptive}C plots the probability distribution $P(q_{6})$, which is the most sensitive measure of the local order
parameters.


\subsection{Orientational correlation function}
\label{or}

\begin{figure}
\centering \resizebox{7cm}{!}
{\includegraphics{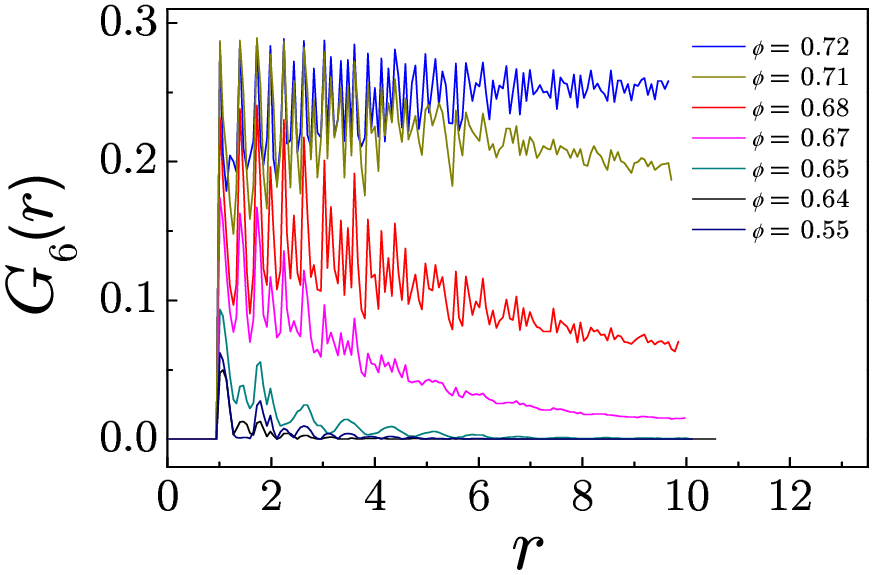}}
\caption{The orientational correlation function Eq. (\ref{G}) for
  different packings with $\phi_j$. $G_6(r)$ vanishes at large $r$
  below $\phi_j=0.64$, while it approaches a nonzero constant when
  $\phi_j>0.64$. The nonzero asymptote of $G_6(r)$ is a signature of
  long-range correlation of orientational order. RCP is a well defined
  singularity at $\phi_j=0.64$ where the orientational symmetry
  breaking occurs.}
\label{orientationalCorrelationFunction}
\end{figure}

The bond-angle correlation functions
(Fig. \ref{orientationalCorrelationFunction}) can be obtained via
\cite{nelson}:

\begin{equation}
G_l\left(r\right)=\frac{4\pi}{2l+1}\sum_{m=-l}^{l}\langle
Q_{lm}\left(\vec{r}\right)Q_{lm}(\vec{0})\rangle,
\label{G}
\end{equation}
where the angular bracket indicates an average over all particles
separated by $\vec{r}$. A non-zero asymptotic value of
$G_6\left(r\right)$ implies a long-range correlation in the
orientational order.  From Fig. \ref{orientationalCorrelationFunction},
it is clear that the crystal phase has long-range orientational order,
while no order can be found in the random phase.

\subsection{Local orientational disorder}
\label{local}

We have established that crystalline structures appear in the
coexistence region and solid-like branch. The remaining question
is what kind of lattice structure dominates in the crystallized
packings. Since the differences of the orientational order
parameters, especially $Q_{6}$, are not significant between
different perfect lattice clusters (such as the icosahedral, FCC,
HCP, BCC, and SC clusters, see \cite{nelson}, Fig. 2) we apply
another measure, the local orientational disorder, to identify the
crystalline nuclei as defined in M. Bargiel and E. M. Tory, {\it Adv. Powder Technol.} {\bf 12}, 533 (2001).

For a given sphere $i$, let
$\theta_{ijk}$ be the angle between the $j$th and $k$th
neighbors. Furthermore let $\theta_{jk}^{\rm fcc}$, $\theta_{jk}^{\rm hcp}$,
$\theta_{jk}^{\rm icos}$ be similarly calculated angles for the perfect
13-sphere fragments of FCC, HCP and icosahedral packings,
$\theta_{jk}^{\rm bcc}$ be the angles for the 8-sphere fragment of a
perfect BCC packing, and $\theta_{jk}^{\rm sc}$ be the angles for the
6-sphere fragment of a perfect SC packing. The local disorders are
defined as following:

\begin{equation}
\theta_{i}^{\rm fcc}=\sqrt{\frac{1}{66}\sum_{j=1}^{11}\sum_{k=j+1}^{12}\left(\theta_{ijk}-\theta_{jk}^{\rm fcc}\right)^2},
\label{fcc}
\end{equation}

\begin{equation}
\theta_{i}^{\rm hcp}=\sqrt{\frac{1}{66}\sum_{j=1}^{11}\sum_{k=j+1}^{12}\left(\theta_{ijk}-\theta_{jk}^{\rm hcp}\right)^2},
\label{hcp}
\end{equation}

\begin{equation}
\theta_{i}^{\rm icos}=\sqrt{\frac{1}{66}\sum_{j=1}^{11}\sum_{k=j+1}^{12}\left(\theta_{ijk}-\theta_{jk}^{\rm icos}\right)^2},
\label{icos}
\end{equation}

\begin{equation}
\theta_{i}^{\rm bcc}=\sqrt{\frac{1}{28}\sum_{j=1}^{7}\sum_{k=j+1}^{8}\left(\theta_{ijk}-\theta_{jk}^{\rm bcc}\right)^2},
\label{bcc}
\end{equation}

\begin{equation}
\theta_{i}^{\rm sc}=\sqrt{\frac{1}{15}\sum_{j=1}^{5}\sum_{k=j+1}^{6}\left(\theta_{ijk}-\theta_{jk}^{\rm sc}\right)^2}.
\label{sc}
\end{equation}

Note that to calculate the values properly, we first need to sort the
angles $\theta_{ijk}$, $\theta_{jk}^{\rm fcc}$, $\theta_{jk}^{\rm
  hcp}$, $\theta_{jk}^{\rm icos}$, $\theta_{jk}^{\rm bcc}$ and
$\theta_{jk}^{\rm sc}$, and then compare them one by one. Also note
that the BCC and SC clusters have fewer neighbors than FCC, HCP and
icosahedral clusters.

\begin{figure}
  \hbox{ {\bf (A)}
    \resizebox{7cm}{!}{\includegraphics{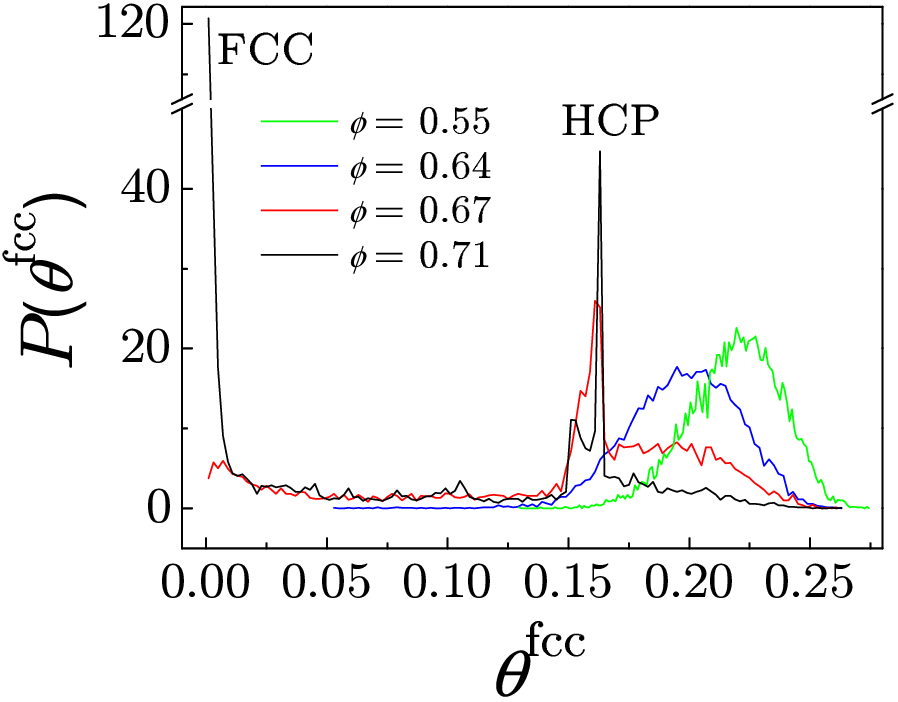}}
    {\bf (B)}
    \resizebox{7cm}{!}{\includegraphics{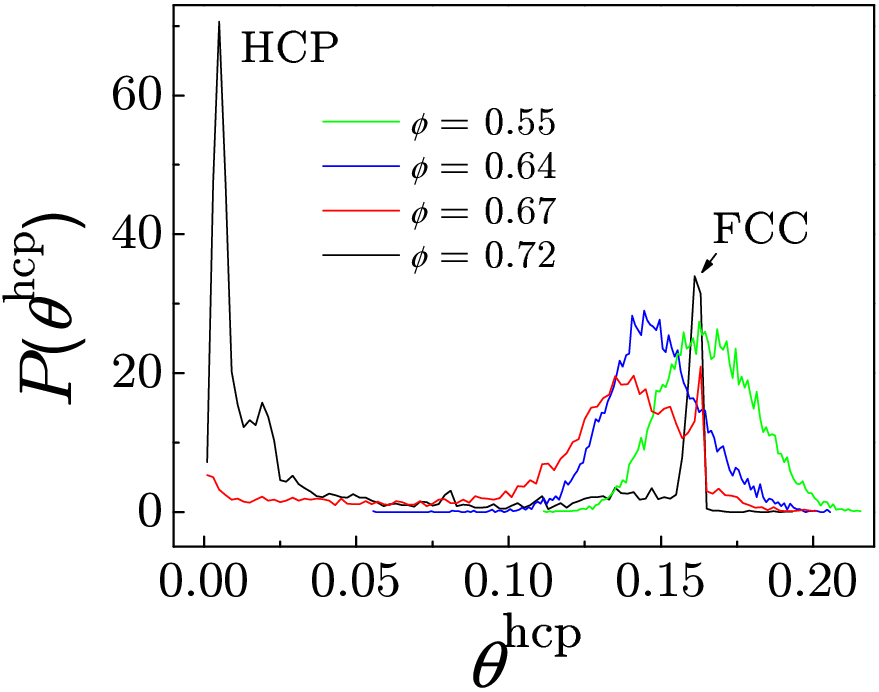}}
  }
  \caption{ {\bf (A)} Distributions of local disorder $\theta_i^{\rm
      fcc}$ defined in Eq. (\ref{fcc}) for packings with different
    $\phi_j$ as indicated. The distribution functions below
    $\phi_j=0.64$ are Gaussian, the center of the peak moves to
    $\theta_i^{\rm fcc}=0.2$ as the volume fraction $\phi_j$
    approaches 0.64. This disordered peak decreases above
    $\phi_j=0.64$ and seems to disappear above $\phi_j=0.68$. Two
    ordered peaks appear after RCP, the one at $\theta_i^{\rm fcc}=0$
    corresponds to FCC structure, while the other one at
    $\theta_i^{\rm fcc}=0.16$ corresponds to HCP. The peak at zero
    eventually evolves to a delta function as the packing structure
    approaches a perfect FCC. We use the median of the FCC peak and the
    disordered peak at $\phi_j=0.64$ as a cutoff to identify local FCC
    structure, ie., particles with $\theta_i^{\rm fcc}<\theta_c^{\rm
      fcc}$ are defined as FCC crystalline nuclei, where
    $\theta_c^{\rm fcc}=0.1$. {\bf (B)} Distributions of local
    disorder $\theta_i^{\rm hcp}$ defined in Eq. (\ref{hcp}) for
    packings with different $\phi_j$ as indicated. The distribution
    functions have similar behavior as those of $\theta_i^{\rm
      fcc}$. The cutoff $\theta_c^{\rm hcp}=0.075$ is used to identify
    HCP crystalline nuclei.} \label{local_disorder}
\end{figure}


Since the $\theta_{i}$'s measure the local disorder in the
packings compared to a particular lattice structure, a perfect
lattice cluster would have a zero value of $\theta$. Any packing
with significant amount of certain lattice clusters would indicate
a peak centered around the origin in the distribution function of
the local disorder $\theta_{i}$ corresponding to that particular
lattice.

The distribution of FCC clusters $P(\theta_i^{\rm fcc})$ for
packings with different $\phi_j$ is shown in Fig.
\ref{local_disorder}A. We find that FCC and HCP dominate in the
crystalline packings for $\phi_j\ge\phi_{\rm melt}$. Indeed we
observe two prominent peaks in the distribution, one at FCC
$\theta_i^{\rm fcc}=0$ and the other at HCP $\theta_i^{\rm
fcc}=0.16$, while BCC, SC and icosahedral ordering are negligible.
The FCC peak dominance indicates that the majority of the clusters
are FCC with a small proportion of HCP clusters.  Similar to the
distributions of local orientational orders shown in Fig.
\ref{descriptive}C, we find no crystalline clusters in the random
packings as evidenced by the Gaussian distributions of
$P(\theta_i^{\rm fcc})$ for $\phi_j\le\phi_{\rm rcp}$ as seen in
Fig.  \ref{local_disorder}A. In the coexistence region, the
distributions are formed by a linear combination of different
phases at melting and freezing. The distribution of HCP clusters
$P(\theta_i^{\rm hcp})$ has similar behavior as $P(\theta_i^{\rm
fcc})$, as showed in Fig.  \ref{local_disorder}B.



\subsection{Cluster analysis of crystalline regions and correlation length}
\label{length}

We are able to define crystalline or nearly crystalline clusters
in a packing based on the local orientational disorder and
mechanical contacts, and visualize them in a 3d plot. The clusters
are defined as follows: First, each node in the clusters is a
sphere with $\theta_i^{\rm fcc}<\theta_c^{\rm fcc}$ or $\theta_i^{\rm hcp}<\theta_c^{\rm hcp}$, where $\theta_c^{\rm fcc}=0.1$ and $\theta_c^{\rm
hcp}=0.075$, as determined in Fig.  \ref{local_disorder}. The definition ensures that the first peak in
$P(\theta_i^{\rm fcc})$ at zero in the distribution function of
Fig. \ref{local_disorder}A is included in this consideration (as
well as the analogous analysis for HCP). Next, if any two nodes
are in mechanical contact, we build a link between the two nodes.
Then the crystalline clusters are those nodes that are linked
together.
The clusters are visualized in Fig. \ref{phase}.

Based on the definition of the crystalline clusters, the size of the
largest cluster in the system and the correlation length of the clusters, $\xi$,
are measured near the melting point. To calculate the correlation
length, we first introduce the radius of gyration,
$R_g\left(s\right)$, of a cluster consisting of $s$ particles:
\begin{equation}
R_g^2\left(s\right)=\frac{1}{2s^2}\sum_{i,j}\left(\vec{r_i}-\vec{r_j}\right)^2,
\end{equation}
then the correlation length is given by
\begin{equation}
  \xi^2=\frac{2\displaystyle\sum_sR_g^2\left(s\right)s^2n_s}{\displaystyle\sum_ss^2n_s},
\end{equation}
where $n_s$ is the number of clusters of size $s$ in the packing.

\begin{figure}
    \resizebox{7cm}{!}{\includegraphics{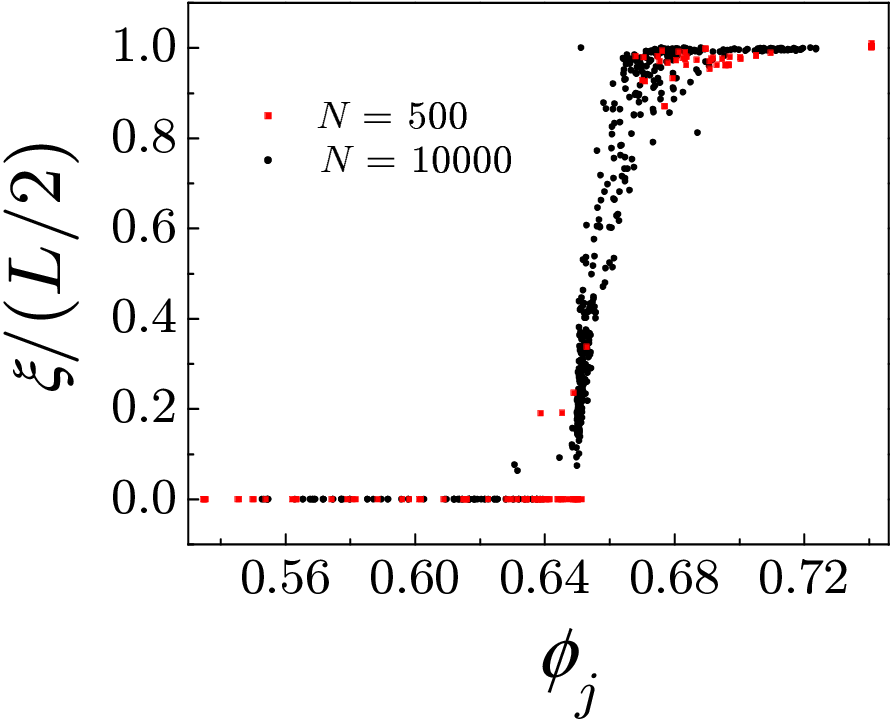}}
    \caption{Correlation length $\xi$ of crystalline clusters. To
      calculate the correlation length, we first introduce the radius
      of gyration, $R_g\left(s\right)$, of a cluster consisting of $s$
      particles. The correlation length of a perfect FCC lattice with
      periodic boundary condition is $L/2$, where $L$ is the system
      size, so the correlation length is scaled by $L/2$ in the
      figure. The crystalline clusters start to percolate at
      $\phi_j=0.68$ as $\xi/(L/2)$ reaches a plateau with value 1. The
      results also show that the percolation at $\phi_{\rm melt}$ does
      not depend on the system size.  } \label{xi}
\end{figure}

The correlation length $\xi$ of the crystalline clusters is
measured near the melting point. Figure \ref{xi} confirms the
linear increase of the size of crystals along the coexistence
region from the freezing point where $\xi=0$ to the melting point.
When the system melts at $\phi_{\rm melt}$, $\xi$ reaches a
plateau consistent with the system size.

\subsection{Radial distribution function}
\label{radial}

\begin{figure}
\centering \resizebox{8cm}{!}
{\includegraphics{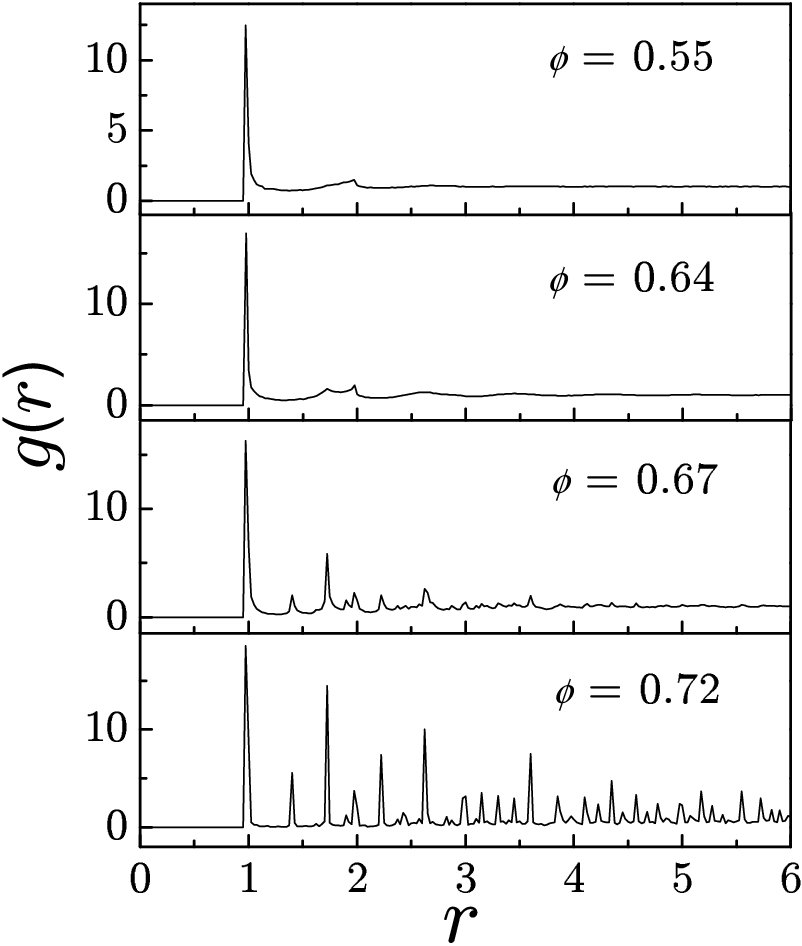}}
\caption{Radial distribution functions $g(r)$ with different volume
  fractions $\phi_j$. The figure clearly show the increasing of long
  range spatial order above $\phi_j=0.64$. }
\label{radialDistributionFunction}
\end{figure}

The radial distribution function $g(r)$ of the packing with volume
fraction 0.72 in Fig. \ref{radialDistributionFunction} shows all the peaks in FCC and HCP packings, which
are indications of long range spatial order. On the other hand,
the radial distribution functions of random packings only have two or three peaks (the second peak splits at $\phi_j=0.64$), corresponding to short
range order. In the coexistence region, $g(r)$ has more peaks than those of random packings, but the magnitude of the peaks decays very fast as the distance $r$ becomes larger. The long-range order increases with the volume
fraction in the coexistence region.

\section{Thermodynamic viewpoint of the RCP transition }
\label{thermo}

\subsection{Free energy}
\label{free}

In the thermodynamics of jammed matter, the internal energy $U$ is
replaced by the volume $W$ (usually called the volume function)
\cite{sirsam}. Other thermodynamic potentials, such as enthalpy $H$,
free energy $F$, Gibbs free energy $G$, are related to the volume
function $W$ as

\begin{eqnarray}
H=W, \\
F=W-XS, \\
G=F=W-XS,
\end{eqnarray}
where $X$ is the compactivity of the system and $S$ the entropy, related as:

\begin{equation}
  \frac{1}{X} = \frac{\partial S}{\partial W}.
\label{x3}
\end{equation}

Since thermodynamic pressure is not considered in our case, the
volume $W$ and enthalpy $H$ are equivalent and the Helmholtz free
energy $F$ and Gibbs free energy $G$ are identical, so that we refer
to them only as the free energy. The differentials of the
thermodynamic potentials are

\begin{eqnarray}
dW = X dS + \mu dN,\\
dF = -S dX + \mu dN,
\end{eqnarray}
where $\mu$ is the chemical potential and $N$ is the number of
particles.


The free energy $F$, or Gibbs free energy $G$ is equal to the
product of $N$ and $\mu$,
\begin{equation}
F=G=N\mu.
\label{mu}
\end{equation}
The free energy and chemical potential are continuous in the first
order phase transition.  On the other hand, the derivative of the
chemical potential, $S=-\frac{\partial \mu}{\partial X}$, is
discontinuous in a first order phase transition.

The compactivity $X$, as well as the entropy density $s=S/N$, can be
calculated from the fluctuations of the Voronoi volumes, which is an
analogy of the Edwards theory to the standard Boltzmann statistical
mechanics.  The definition of a Voronoi cell is a convex polygon whose
interior consists of all points closer to a given particle than to any
other.  The Voronoi volume of a given particle, $V_{\rm vor}$, is the
volume of such a Voronoi cell (see \cite{song} for more details).

We define the Voronoi fluctuations as $\sigma^2 \equiv \langle
w^2_{\rm vor} \rangle - \langle w_{\rm vor} \rangle^2$, where $w_{\rm
  vor} = V_{\rm vor}/V_g$ is the reduced Voronoi volume of each
particle and the average is done over all the particles in a packing.
The quantities $\sigma_{1}(\phi_j)$ and $\sigma_{2}(\phi_j)$ denote
the Voronoi fluctuations in the disordered and ordered phases,
respectively, as a function of $\phi_j$.

We use the Einstein fluctuation theorem for jammed matter which is
obtained from the similar relation in equilibrium systems by replacing
the energy by the volume fluctuations \cite{nowak,swinney,chris}:

\begin{equation}
\langle (\delta W)^2\rangle = k_B X^2 \frac{\partial W}{\partial X}.
\label{delta}
\end{equation}
Here we assume that $k_B$ plays the role of the Boltzmann constant
in thermodynamics. Its value could be set to unity without
changing the obtained results since in this context it just
defines the units of entropy (in the main text we set $k_B=1$ to
simplify). This means that we measure the compactivity in units of
volume $V_g$ and that the entropy, which has units of $k_B$, is
dimensionless.  In terms of the Voronoi fluctuation $\sigma$ and
volume fraction $\phi_j$, Eq. (\ref{delta}) reads:

\begin{equation}
  \sigma_i^2=- \frac{k_B}{V_g} \Big(\frac{ X}{\phi_j}\Big)^2 \frac{\partial \phi_j}{\partial X}, \,\,\,\ i=1,2.
  \label{sigma}
\end{equation}

Equation (\ref{sigma}) applies to the disordered and ordered branches,
$\sigma_1$ and $\sigma_2$, separately.  It does not apply to the
coexistence region, since it requires the pure phases to calculate
the fluctuations.

Using Eq. (\ref{sigma}) we can calculate the thermodynamic quantities
by integration.
We first obtain the compactivity from:

\begin{subequations}
\begin{align}
  \frac{1}{X(\phi_j)}= \frac{k_B}{V_g} \int_{\phi_{\rm rlp}}^{\phi_j}
  \frac{d\phi}{\phi^2\sigma_{1}^{2}(\phi)}+\frac{1}{X_{\rm rlp}},
  \,\,\, \,\,\,\,\, \phi_{\rm rlp} \leq \phi_j \leq \phi_{\rm rcp},
  \label{x2a}\\
  \frac{1}{X(\phi_j)}= \frac{k_B}{V_g} \int_{\phi_{\rm
      melt}}^{\phi_j}\frac{d\phi}{\phi^2\sigma_{2}^{2}(\phi)}+\frac{1}{X_{\rm
      melt}}, \,\,\,\,\,\,\,\, \phi_{\rm melt} \leq \phi_j \leq
  \phi_{\rm fcc},
  \label{x2b}
\end{align}
\label{x2}
\end{subequations}
where $X_{\rm melt}=X(\phi_{\rm melt})$ is the compactivity of the
packing at the melting point and $X_{\rm rlp}=X(\phi_{\rm rlp})$ at
RLP.


The entropy density is then obtained by a second integration using the
definition Eq. (\ref{x3}) which reads:

\begin{equation}
  \frac{1}{X} = -\frac{\phi_j^2}{V_g} \frac{\partial s}{\partial \phi_j}.
\label{x4}
\end{equation}
 We integrate Eq. (\ref{x4}) for each branch to obtain:

\begin{subequations}
\begin{align}
  s(\phi_j)=s_{\rm rcp}+V_g \int_{\phi_j}^{\phi_{\rm rcp}}
  \frac{d\phi}{X(\phi)\phi^{2}}, \,\,\,\,\,\,\,\, \phi_{\rm rlp} \leq
  \phi_j \leq \phi_{\rm rcp},
\label{s2a}  \\
  s(\phi_j)= V_g \int_{\phi_j}^{\phi_{\rm fcc}}
  \frac{d\phi}{X(\phi)\phi^{2}}, \,\,\,\,\,\,\,\, \phi_{\rm melt} \leq
  \phi_j \leq \phi_{\rm fcc}.
\label{s2b}
\end{align}
\label{s2}
\end{subequations}
The entropy of FCC, $s_{\rm fcc}$, is zero in the thermodynamic limit.

Equations (\ref{x2}) and (\ref{s2}) require three constants of
integration: $X_{\rm rlp}, X_{\rm melt}$ and the entropy of RCP:
$s_{\rm rcp}$.  We now introduce three extra constraints to close the
system.  First, there are two conditions for equilibrium between two phases
in jammed matter \cite{debenedetti,sirsam}: (a) ``thermal'' equilibrium
\begin{equation}
\label{xc}
X_{\rm melt}=X_{\rm rcp}\equiv X_c,
\end{equation}
where $X_c$ is the critical compactivity at the transition. (b) The
equality of the chemical potentials of the two phases at the melting
and the freezing RCP point: $\mu_{\rm melt}=\mu_{\rm freez}$ due to
the conservation of number of particles. This is analogous to the
equality of the free energy density at the transition from
Eq. (\ref{mu}), which allows to calculate the entropy at the freezing
point via
$f_{\rm melt}=f_{\rm rcp}$:
\begin{equation}
  s_{\rm rcp}=s_{\rm melt}+V_g \Big(\frac{\omega_{\rm rcp}-\omega_{\rm melt}}{X_{c}}\Big),
\label{srcp}
\end{equation}
and the entropy of fusion is then:
\begin{equation}
  \Delta s_{\rm fus} \equiv s_{\rm rcp}-s_{\rm melt}= V_g
  \Big(\frac{\omega_{\rm rcp}-\omega_{\rm melt}}{X_{c}}\Big).
\end{equation}


The final constant of integration to be obtained is the compactivity
at the RLP point.  As a first order approximation, $X_{\rm rlp}$ can
be taken as infinite as has been shown in previous experimental and
numerical studies \cite{swinney,chris}.  However, when we compare the
obtained entropy Eq. (\ref{s2}) with an independent measure of the
entropy using Shannon information theory (explained below) we find
that there are slightly differences between both values of the
entropy. Therefore, we consider $X_{\rm rlp}$ as a fitting parameter
to be obtained by fitting the result of Eq. (\ref{s2}) with the
Shannon entropy which in principle does not require any integration
constant to be calculated. We note that using the fitted value of
$X_{\rm rlp}$ instead of infinity does not change the final results,
specifically the values of $X_c$ and the entropy of fusion, even
though the obtained $X_{\rm rlp}$ is ``far'' from infinite.

We summarize the calculation as follows: (a) We assume a value of
$X_{\rm rlp}$ (which is later fitted with Shannon entropy) and
integrate Eq. (\ref{x2a}) from RLP to $\phi_{\rm rcp}$ and obtain the
compactivity $X_{\rm rcp}=X(\phi_{\rm rcp})$.  (b) Using
Eq. (\ref{xc}) we obtain $X_{\rm melt}$ (or $X_c)$.  (c) Using $X_{\rm
  melt}$, we integrate Eq. (\ref{x2b}) to obtain the $X(\phi_j)$ in
the ordered branch, thus completing the calculation of the
compactivity equation of state.  (d) We integrate Eq. (\ref{s2b}) up
to $\phi_{\rm melt}$ to obtain the entropy of the melting point:
$s_{\rm melt}$.  (e) Using Eq. (\ref{srcp}) we obtain the entropy at
RCP, $s_{\rm rcp}$.  (f) This value is then plugged into
Eq. (\ref{s2a}) to finish the calculation of $s(\phi_j)$ by a final
integration.  The above procedure is repeated for different values of
$X_{\rm rlp}$ starting from infinite up to a finite value that will
match the Shannon entropy as discussed below.


\subsection{Shannon entropy}
\label{en}

Entropy of jammed matter can be calculated in two different ways: {\it
  (i)} The entropy from fluctuation theory explained above.  {\it
  (ii)} The entropy from information theory \cite{chris}, so called
''Shannon entropy", related to configurational disorder since
topologically equivalent structures are considered as the same
state. Shannon entropy attempts to measure the disorder in a
string of information as defined in the seminal work of Shannon.
This concept has been adapted to the measurement of the
configurational entropy in physical systems defined through a
contact network by Vink and Barkema [R. L. C. Vink, G. T. Barkema,
{\it Phys. Rev. Lett.} {\bf 89}, 076405 (2002)]. In this work it
was shown that the entropy obtained from the thermodynamic
integration of fluctuations (or equivalently the specific heat)
and the Shannon entropy are equivalent and accurately describe
amorphous silicon and vitreous silica networks.  The method has
been extended to calculate the entropy of packings of granular
materials in \cite{chris}.
Below we explain the main details.

The advantage of the Shannon entropy calculation over the
thermodynamic integration is that it does not require a constant of
integration as in Eq. (\ref{s2}).  For each state of jammed matter, we
associate a probability of occurrence $p_{i}$ to the state, which is
calculated as follows.

We use the Voronoi cell and Delaunay triangulation for each
particle to define a Voronoi network by considering contacts when
a Voronoi side is shared between two particles, and hence are
Delaunay contacts. A graph is constructed as a cluster of $n$
particles that are Delaunay contacts, and by means of graph
automorphism [B. D. McKay, Nauty user's guide (version 1.5), Tech.
Rep. TR-CS-90-02, Australian National University (1990)] can be
transformed into a standard form or ``class" $i$ of topologically
equivalent graphs with a probability
of occurrence $p(i)$.  In practice, we determine $p(i)$ by
extracting a large number $m$ of clusters of size $n$ from the
system and count the number of times, $f_i$, a cluster $i$ is
observed, such that:

\begin{equation}
p(i)=f_i/m. \label{prob}
\end{equation}

Then the Shannon entropy is defined as:
\begin{equation}
  H\left(n\right)=-\sum p_{i}\ln p_{i},
\end{equation}
where we have again assumed the ``Boltzmann-like" constant in
front to be one. The Shannon entropy density is obtained  as:
\begin{equation}
  s_{\rm shan}=\lim_{n\rightarrow\infty}\left[
    H\left(n+1\right)-H\left(n\right)\right],
\label{sh}
\end{equation}
by linear fitting of the extensive part of the Shannon entropy.

In general, the fluctuation entropy from Eq. (\ref{s2}) is greater
than the Shannon entropy from Eq. (\ref{sh}) because Shannon entropy
only counts configurational disorder and additional entropy could
arise from freedom to move grains within the clusters of $n$ particles
without disrupting the Delaunay network.  However, we discover that
the fluctuation entropy obtained from Eq. (\ref{s2}) and Shannon
entropy from Eq. (\ref{sh}) only differ by a scaling constant $k=0.1$,
ie., $k s=s_{\rm shan}$, see Fig.  \ref{fluctuations}B.  Beyond this
multiplicative constant the agreement between both estimations of the
entropy is very good. Due to finite size effects the Shannon entropy
also gives a nonzero value of the entropy of FCC, $s_{\rm
  fcc}=0.6$. This value is subtracted from the calculations.  By
fitting $s_{\rm shan}$ with the thermodynamic entropy we obtain the
final constant of integration in Eq. (\ref{x2a}), $X_{\rm rlp}= 0.5
V_g$. While it is obvious that this constant is far from the infinite
value which is expected and used in \cite{swinney,chris} for the
RLP limit, we notice that our final results are not very sensitive to
the exact value of $X_{\rm rlp}$. For instance, by setting $X_{\rm
  rlp}\to \infty$ the fitting of the fluctuation entropy is slightly
off in comparison with $s_{\rm shan}$ only in the vicinity of RLP but
the values of $X_c$ and the entropy of fusion do not have appreciable
change.

\begin{figure}
  \centering \resizebox{8cm}{!}  {\includegraphics{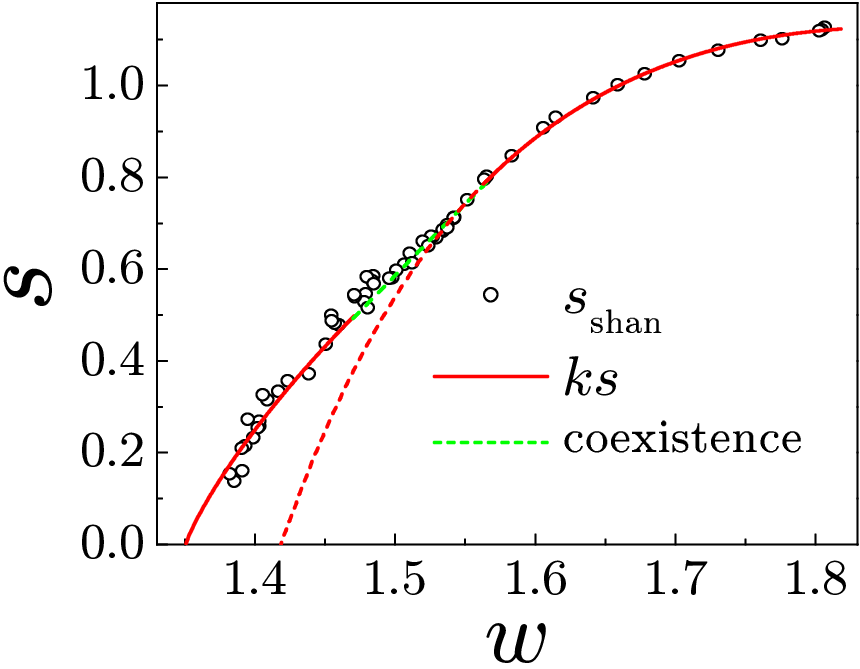}}
\caption{
  Entropy versus reduced volume function $\omega=1/\phi_j$. }
  \label{entropy2}
\end{figure}

Figure \ref{compactivity} in the main text shows the thermodynamic
quantities in the first order phase transition of jammed matter. They
are consistent with the general thermodynamic picture.
Figure \ref{entropy2} plots the entropy versus $w$ showing a linear
dependence in the coexistence region.  The entropy is an interpolation
of the form: $s_x = x s_{\rm melt} + (1-x) s_{\rm freez}$ where $x$ is
the concentration of crystal clusters in the coexistence. Since
$X=\partial W/\partial S$, the linearity of $s$ between 0.64 and 0.68
is a manifestation of the coexistence of two phases at a constant
$X_c$.

\end{document}